\documentclass[a4paper,fleqn]{cas-sc}

\usepackage[numbers]{natbib}
\usepackage[boxed,commentsnumbered,ruled,linesnumbered]{algorithm2e}

\newtheorem{definition}{Definition}  %gws
\newtheorem{property}{Property}   % gws
\newtheorem{strategy}{Strategy}    % gws
\begin{document}

\shorttitle{Targeted Mining of Top-$k$ High Utility Itemsets} 

\shortauthors{Shan Huang et~al.}

\title [mode = title]{Targeted Mining of Top-$k$ High Utility Itemsets}

\author[1]{Shan Huang}
\ead{shuang9901@gmail.com}
\address[1]{College of Cyber Security, Jinan University, Guangzhou 510632, China}

\author[1,2]{Wensheng Gan}
\cormark[1]
\ead{wsgan001@gmail.com}
\address[2]{Pazhou Lab, Guangzhou 510330, China}
\cortext[cor1]{Corresponding author}

\author[1]{Jinbao Miao}
\ead{osjbmiao@gmail.com}

\author[3]{Xuming Han}
\ead{hanxuming@jnu.edu.cn}
\address[3]{College of Information Science and Technology, Jinan University, Guangzhou 510632, China}

\author[4]{Philippe Fournier-Viger}
\ead{philfv@qq.com}
\address[4]{College of Computer Science and Software Engineering, Shenzhen University, Shenzhen 518060, China}

\begin{abstract}
	Finding high-importance patterns in data is an emerging data mining task known as High-utility itemset mining (HUIM). Given a minimum utility threshold, a HUIM algorithm extracts all the high-utility itemsets (HUIs) whose utility values are not less than the threshold. This can reveal a wealth of useful information, but the precise needs of users are not well taken into account. In particular, users often want to focus on patterns that have some specific items rather than find all patterns. To overcome that difficulty, targeted mining has emerged, focusing on user preferences, but only preliminary work has been conducted. For example, the targeted high-utility itemset querying algorithm (TargetUM) was proposed, which uses a lexicographic tree to query itemsets containing a target pattern. However, selecting the minimum utility threshold is difficult when the user is not familiar with the processed database. As a solution, this paper formulates the task of targeted mining of the top-$k$ high-utility itemsets and proposes an efficient algorithm called TMKU based on the TargetUM algorithm to discover the top-$k$ target high-utility itemsets (top-$k$ THUIs). At the same time, several pruning strategies are used to reduce the memory consumption and execution time. Extensive experiments show that the proposed TMKU algorithm has good performance on real and synthetic datasets.
\end{abstract}


\maketitle

%%%%%%%%%%%%%%%%%%%%%%%%%%%%%%%%%%%%%%%
\section{Introduction} \label{sec:introduction}

As digital systems are widely used, the era of big data has arrived, with massive amounts of data being generated. Therefore, figuring out how to mine useful information from complex data has become a major issue. Data mining software assists clients in finding correlations in millions or billions of records, allowing decision-makers to make more informed decisions faster \cite{fournier2022pattern,gan2019utility,gan2019survey}. This technology is influencing many areas of daily life (e.g., market analysis \cite{hemalatha2012market}, biomedicine \cite{peek2014technical}, and network security management \cite{chai2009analyzes}). A typical application is market basket analysis \cite{kuriakose2017efficient}. Market basket analysis primarily analyzes the consumption records of customers to uncover important correlations between products and services, allowing them to rearrange products on shelves or design a combination of promotional packages that users are interested in. For example, if a supermarket's sales records show that users often buy bread and milk at the same time, bundling them on the same shelf can increase sales and revenue.

Previous studies have proposed numerous frequent itemset mining algorithms (FIM) \cite{aggarwal2014frequent, kuriakose2017efficient}, among which the most famous are Apriori \cite{agrawal1994fast} and FP-growth \cite{han2000mining}. Traditional frequent pattern mining algorithms use only one metric, support, to find patterns that users are interested in. As a result of applying the frequent itemset model, itemsets are often found to have a high selling frequency but yield a low profit. But some luxury goods that are not widely sold are highly profitable and deserve attention. To address this issue, utility-oriented pattern mining (UPM) \cite{gan2020huopm,krishnamoorthy2015pruning, song2014mining,wu2021haop} algorithms have been proposed. Internal and external utilities are both included in the concept of utility. The internal utility usually refers to how many times a good appears, while the profit per unit generated by each good is known as the external utility. UPM has been extensively studied in recent decades \cite{gan2018survey,gan2021survey}, since it is beneficial to consider the profitability and relative importance of each item. Some representative UPM algorithms are HUI-Miner \cite{liu2012mining}, which is based on the utility-list structure; UP-Growth \cite{tseng2010up} based on the UP-tree structure; and EFIM \cite{zida2015efim} based on data projection.
 
Although high-utility itemset mining algorithms are able to collect some useful information in specific practical applications, they are not suitable for target-oriented mining tasks. The targeted high-utility itemset querying (TargetUM) algorithm \cite{miao2021targeted} was introduced to resolve this issue by giving consideration to user needs. Take consumer purchases as an example: people pay more attention to the items on their shopping list and less to other items. Therefore, retailers can offer different product recommendations according to the various needs of users as well as discounts on goods in stock. Hence, target-oriented high-utility itemset mining (THUIM) can filter out many irrelevant patterns in the data to precisely focus on those containing items of interest. However, it is complex and laborious to select the minimum utility threshold in these utility-driven data mining algorithms. In particular, if a user is not familiar with a database, he will not have a good understanding of the utilities of itemsets and the content of the transactions. When the utility threshold is set too high, only a small percentage of target high-utility patterns are captured, while a lot of irrelevant information will be generated for a low utility threshold, which has no value to the user. As a result, depending on the number of items and utility distribution in each database, different utility thresholds must be set. In general, decision makers will always repeatedly test and modify the utility threshold until they find a threshold value that will produce the appropriate number of target high-utility itemsets, which can take a long time and is inconvenient.

Consequently, the problem of targeted mining of top-$k$ high utility itemsets is formulated in this paper, for which the user can directly set the number of itemsets to be mined. An algorithm called TMKU (Targeted Mining of top-$k$ high Utility itemset) is designed to perform this task efficiently. TMKU has proved effective and efficient in experiments. The main contributions of this study are outlined below:

\begin{itemize}
	\item This paper considers two important aspects: utility mining and targeted pattern discovery. As far as we know, the TMKU algorithm is the first algorithm to incorporate the concept of top-$k$ mining into target-based utility mining.

	\item A trie structure is used to query the target itemsets and a new data structure, namely TopKMap is used to efficiently store and retrieve the top-$k$ target itemsets, which is beneficial to reduce the query time and search space.

	\item The TMKU algorithm utilizes several upper bounds. Three pruning strategies are used to reduce the search space. In addition, two target utility raising strategies are adopted to quickly raise the threshold and ensure that all top-$k$ THUIs are discovered.

	\item Extensive experiments were conducted on both real and synthetic datasets. The results demonstrate that TKUM is efficient and effective.	
\end{itemize}

The rest of this paper is composed of six sections. A brief review of related work is provided in Section \ref{sec:relatedwork}. Section \ref{sec:preliminaries} describes the top-$k$ THUIM problem and related definitions. Section \ref{sec:algorithm} introduces the proposed TMKU algorithm, as well as the main strategies and techniques applied. The experimental results are presented and evaluated in Section \ref{sec:experiment}. Ultimately, Section \ref{sec:conclusion} concludes this paper and provides an outlook on promising research directions.

\section{Related Work} \label{sec:relatedwork}

The section introduces the research related to the paper, which can be divided into three parts: (1) high-utility pattern mining; (2) top-$k$ utility itemset mining; and (3) targeted pattern mining.

\subsection{High-utility pattern mining}

Frequent itemset mining (FIM) \cite{aggarwal2014frequent,agrawal1994fast,han2000mining} has been extensively studied for decades. However, relying only on frequency cannot bring enough benefits to users. Factors such as quantity and profit should also be considered. For this reason, Chen \textit{et al.} \cite{chan2003mining} put forward a new task called high-utility itemset mining (HUIM). Since then, utility mining research has developed rapidly \cite{gan2021survey,lin2016efficient,song2016high,wu2021haop}. For the convenience of discussion, high-utility itemset mining algorithms are grouped into the following three categories:

\textbf{Apriori-based algorithms}: Since Agrawal \textit{et al.} \cite{agrawal1993mining} proposed the Apriori property in 1994, lots of algorithms based on Apriori have been published. For example, Liu \textit{et al.} \cite{liu2005two} introduced the prominent Two-Phase algorithm to handle the difficulty that the utility, unlike frequency, is neither monotone nor anti-monotone. That algorithm uses an overestimation of the utility called \textit{TWU} (Transaction Weighted Utilization) to find candidate itemsets in a first phase. Thereafter, in a second phase, the database is searched again to determine the exact utility value of each candidate itemset. The IIDS algorithm \cite{li2008isolated} is an improved version of Two-Phase that discards isolated items to shrink the search space. However, the common disadvantage of Apriori-like algorithms is that plenty of candidate patterns are generated, resulting in considerable computational costs and memory consumption.

\textbf{Tree-based algorithms}: Tseng \textit{et al.} \cite{tseng2010up} designed the UP-tree structure, a utility-pattern tree, and introduced the UP-Growth algorithm inspired by FP-Growth. Subsequently, other versions of tree-based algorithms \cite{song2014mining,tseng2012efficient} have been presented. In general, utilizing the UP-tree can prevent many meaningless database scans. When working with large-scale databases, however, this structure grows increasingly complex and occupies a massive amount of memory.

\textbf{Other structure-based algorithms}: HUI-Miner \cite{liu2012mining} utilizes a novel data structure known as a utility-list, which avoids the difficulty of generating numerous candidates. Moreover, the FHM algorithm \cite{fournier2014fhm} reduces the cost of join operations by using a tighter upper bound, which results in outperforming HUI-Miner. However, the join operation on lists of these algorithms takes time and memory. Thus, Zida \textit{et al.} \cite{zida2015efim} proposed the EFIM algorithm with high-utility database projection (HDP) and high-utility transaction merging (HTM) techniques to lower the expensive cost of database passes. The utility-list-based CoUPM algorithm for correlated utility-based pattern mining \cite{gan2019correlated}. In summary, these algorithms integrate various strategies to discover HUIs as efficiently as possible.

\subsection{Top-$k$ utility itemset mining}

Although the above algorithms are effective in finding the desired set of itemsets, the efficiency of mining is strongly related to the selection of the minimum utility threshold. However, it is not easy to identify an appropriate threshold. Many top-$k$ pattern mining algorithms were thus designed to directly discover the set of top-$k$ HUIs, rather than asking users to specify a utility threshold. Top-$k$ HUIM algorithms mainly consist of two types: the first is the two-phase algorithms, and the other is the one-phase algorithms.

\textbf{Two-phase algorithms}: The task of discovering the top-$k$ HUIs was proposed by Wu \textit{et al.} \cite{wu2012mining} with the TKU algorithm, which outperformed HUIM algorithms in terms of speed. The TKU algorithm is a two-phase algorithm. In the first phase, a UP-Tree is built, and promising top-$k$ HUIs are generated. Then, in the second phase, the desired top-$k$ HUIs are selected among them. TKU applies several strategies to filter unpromising candidates during the search \cite{tseng2015efficient} and achieve higher efficiency. Subsequently, REPT \cite{ryang2015top} was introduced with optimizations to record and pre-calculate the utility of items to prune the search space effectively and raise the minimum utility threshold. REPT uses a tree structure and pre-evaluation matrixes as tools to store utility information. However, these two-phase algorithms still generate large sets of candidates, which causes unreasonably long runtimes and high memory usage.

\textbf{One-phase algorithms}: For top-$k$ HUIM, the one-phase TKO algorithm \cite{tseng2015efficient} was developed to solve the shortcomings of two-phase algorithms. TKO takes advantage of the utility-list structure of HUI-Miner, and outperforms the TKU and REPT algorithms according to experiments \cite{tseng2015efficient}. Similarly, another one-phase algorithm called KHMC \cite{duong2016efficient} also discovers the top-$k$ HUIs by using the utility-list structure. In KHMC, an estimated utility co-occurrence pruning (EUCP) technique is applied, which is based on precalculating the TWU of 2-itemsets. Moreover, the algorithm also adds another pruning strategy named early abandoning to avoid completely constructing the lists of unpromising itemsets. Three threshold-raising strategies are able to significantly shrink the search space and enhance the algorithm's efficiency. The THUI algorithm \cite{krishnamoorthy2019mining} has better performance thanks to introducing the concept of Leaf Itemset Utility (LIU), a triangular matrix, which can be implemented with only a small amount of memory to store utility information. Besides, the LIU-E and LIU-LB threshold raising strategies also accelerate the mining speed of the algorithm. THUI greatly outperforms TKO and KHMC, especially for dense or large datasets.

In addition, there are various other top-$k$ pattern mining problems and variations, such as mining top-$k$ sequential patterns \cite{zhang2021tkus}, mining top-$k$ HUIs in data streams \cite{cheng2021etkds}, discover top-$k$ high-utility sequential patterns \cite{zhang2021tkus}, and mining top-$k$ HUIs with negative utility values \cite{sun2021mining}.

\subsection{Targeted pattern mining}

Those algorithms listed above are designed to find all itemsets that meet a single predetermined criterion. Target-oriented query algorithms give an alternative solution to this problem by filtering out unnecessary information. Rather than searching for numerous but mostly insignificant items, the user can enter any target and then discover patterns containing the desired items. Several target-oriented query algorithms based on frequency have been developed in earlier studies. These interactive methods are capable of returning results containing a target. Kubat \textit{et al.} \cite{kubat2003itemset} were among the first to address the issue of processing target queries in a transactional database. They implemented target query processing algorithms for association mining by creating itemset trees that can be progressively updated. Fournier-Viger \textit{et al.} \cite{fournier2013meit} developed the Memory Efficient Itemset Tree (MEIT) to further reduce memory requirements. The tree is optimized to perform incremental modifications when new transactions are inserted, and it employs a node-compression method. For multi-objective mining of big data, the guided FP-growth (GFP-growth) algorithm based on FP-Growth was proposed by Shabtay \textit{et al.} \cite{shabtay2018guided}. In particular, many experiments have illustrated the excellent performance of the algorithm on imbalanced data. Target-oriented mining has also been studied and applied to discover sequential patterns. The targeted mining algorithm for sequential patterns proposed by Chueh \textit{et al.} \cite{chueh2010mining} speeds up the search for the target itemsets by using the reversion of the original sequence and comparing the reversed sequence with the related itemsets. Furthermore, clustering analysis is applied to automatically set time partition values for the task of time-interval sequential pattern mining. A novel target-oriented sequential pattern mining approach was presented by Chand \textit{et al.} \cite{chand2012target}, which uses RFM (recency, frequency, and monetary) constraints. As a result, fewer database projections are done, and the space complexity is reduced. To remove some useless or irrelevant patterns in high utility sequential pattern mining \cite{zhang2021shelf,gan2021explainable}, the TUSQ algorithm \cite{zhang2021tusq} first introduced the concept of utility into target sequence queries. The algorithm does not focus on frequency like previous algorithms, but rather on utility. Recently, the TargetUM algorithm \cite{miao2021targeted} has been proposed to fill the gap and perform target-oriented mining in HUIM.

In general, the TargetUM algorithm provides an integrated approach for high-utility mining with a target query, which serves as the foundation for this research. However, there are no studies combining top-$k$ high-utility methods with target pattern queries. This paper introduces the problem of targeted utility mining with the concept of top-$k$ patterns to prevent the generation of large sets of HUIs and to accurately and quickly process target queries.

\section{Preliminaries and Problem Statement}
\label{sec:preliminaries}

This section introduces the redesigned data structure and briefly describes key definitions concerning the problem of targeted top-k high utility itemset mining. $\mathcal{D}$ denotes a transaction database represented as multiple transactions \{$t_1$, $t_2$, $t_3$, $\ldots$, $t_n$\}. Each transaction contains one or more items from a set denoted as \{$i_1$, $i_2$, $i_3$, $\ldots$, $i_m$\}. The transaction identifier of a transaction $t_\textit{id}$ is $id$. In the case where an itemset includes $l$ items, it is called an $l$-itemset. Each item in a transaction can be annotated with a positive integer value indicating a purchase quantity (also called internal utility value or count). For example, $t_1$ = \{$(b,4)$, $(d,3)$, $(e,1)$\} represents four units of an item $b$, three units of an item $d$, and one unit of an item $e$. The itemset $\{ b,d,e\}$ is called a 3-itemset.

\begin{minipage}{\textwidth}
	\begin{minipage}[t]{0.45\textwidth}
		\centering
		\makeatletter\def\@captype{table}\makeatother\setlength{\belowcaptionskip}{10pt}
		\caption{Example database}
		\label{Tab:example} 
		\begin{tabular}{ccc}
			\hline
			$T_{id}$ &   Transaction &  Count    \\ \hline
			$t_1$    & $\{b,d,e\}$    & \{4,3,1\} \\
			$t_2$    & $\{a,b,e\}$  & \{1,3,2\} \\
			$t_3$    & $\{c,d\}$    & \{3,1\} \\
			$t_4$    & $\{a,b,c,f,g\}$    & \{3,1,4,2,1\} \\
			$t_5$    & $\{a,b\}$    &  \{1,2\} \\
			$t_6$    & $\{b,c,d,f\}$     & \{1,3,4,2\} \\
			$t_7$    & $\{a,e,g\}$   &  \{4,1,2\} \\ 
			\hline
			\\
		\end{tabular}
	\end{minipage}
	\begin{minipage}[t]{0.45\textwidth}
		\centering
		\makeatletter\def\@captype{table}\makeatother\setlength{\belowcaptionskip}{10pt}
		\caption{High-utility itemsets w.r.t \textit{minUtil} = \$25}
		\label{Tab:hui}
		\begin{tabular}{cccc}
			\hline
			Itemset &   Utility   & Itemset &   Utility   \\\hline
			$\{f,c,d,b\}$    & \$27  &$\{c,d,b\}$    & \$25 \\
			$\{e,d,b\}$    &  \$26 & $\{d\}$    & \$32\\
			$\{e,b\}$    & \$27  & $\{d,b\}$    & \$43  \\
			$\{c,d\}$    & \$32  &$\{b\}$    & \$33\\			
			\hline
		\end{tabular}
	\end{minipage}
	\begin{minipage}{\textwidth}\vspace{-0.5cm} 
	\end{minipage}
\end{minipage}

\begin{definition}
	\rm (The utility of the itemset) There are two types of utility values in $\mathcal{D}$. On the one hand, in each transaction $t_\textit{id}$, the internal utility of an item $i \in t_\textit{id}$ is denoted as $iu(i,t_\textit{id})$. On the other hand, the external utility of an item $i$ is a positive number denoted as $eu(i)$ that is utilized to indicate its relative weight or importance. Thus, in the transaction $t_\textit{id}$, $u(i,t_\textit{id}$) represents the utility of item $i$, which is the product of $eu(i)$ and $iu(i,t_\textit{id})$, that is $u(i,t_\textit{id})$ = $eu(i)$  $\times$  $iu(i,t_\textit{id})$). For an itemset $X$ and a transaction $t_\textit{id}$ containing $X$  ($X \subseteq t_\textit{id}$), the utility of $X$ in $t_\textit{id}$ is calculated as  $u(X,t_\textit{id})$ = $\sum_{i_{\ell}\in X \wedge X \subseteq t_\textit{id}}u(i_{\ell},~ t_\textit{id})$.  The notation $u(X)$ denotes the utility of $X$ in $\mathcal{D}$ and it is calculated as the sum of the utility of $X$ in all transactions containing $X$, that is $u(X)$ = $\sum_{X\subseteq t_{id} \in \mathcal{D}}{u(x,t_{id})}$.
\end{definition}

For example, the external utility of all items from the example database are \{($a$: 1), ($b$: 3), ($c$: 2), ($d$: 4), ($e$: 2), ($f$: 1), ($g$: 2)\}. In Table \ref{Tab:example}, $iu(b,t_2)$ = \$3, $eu(b)$ = \$3, then $u(b,t_2)$ = $iu(b,t_2)$ $\times$ $eu(b)$ = 3 $\times$ \$3 = \$9.  $u(\{ae\},t_2)$ = $u(a,t_2)$ + $u(e,t_2)$ = 1 $\times$ \$1 + 2 $\times$ \$2 = \$5, $u(\{abc\},t_4)$ = $u(a,t_4)$ + $u(b,t_4)$ + $u(c,t_4)$ = 3 $\times$ \$1 + 1 $\times$ \$3 + 4 $\times$ \$2 = \$14. Since ${d}$ has appeared in $t_1$, $t_3$, and $t_6$, $u(d)$ = $u(d,t_1)$ + $u(d,t_3)$ + $u(a,t_6)$ = 3 $\times$ \$4 + 1 $\times$ \$4 + 4 $\times$ \$4 = \$32, and $u(\{ab\})$ = $u(\{ ab\},t_2)$ + $u(\{ ab\},t_4)$ + $u(\{ ab\},t_5)$ = $u(a,t_2)$ + $u(b,t_2)$ + $u(a,t_4)$ + $u(b,t_4)$ + $u(a,t_5)$ + $u(6,t_5)$ = 1 $\times$ \$1 + 3 $\times$ \$3 + 3 $\times$ \$1 + 1 $\times$ \$3 + 1 $\times$ \$1 + 2 $\times$ \$3 = \$23.

\begin{definition}
	\rm (The transaction-weighted utility of itemset) The  notation $tu(t_\textit{id})$ refers to the utility of the transaction $t_\textit{id}$ and is defined by the sum of the utilities of all the items in $t_\textit{id}$. In other words, $tu(t_\textit{id}$) = $\sum _{i_{\ell}\in t_\textit{id}}u(i_{\ell},t_\textit{id})$. The transaction-weighted utility of an itemset $X$ in $\mathcal{D}$ is denoted as \textit{TWU(X)} and is the sum of the utilities of $X$ in all the transactions containing $X$ in $\mathcal{D}$, that is \textit{TWU(X)} = $\sum_{X\subseteq t_{id} \wedge t_{id} \in \mathcal{D}}u(X,~ t_\textit{id})$ $ = $ $\sum_{X\subseteq t_\textit{id} \wedge t_\textit{id}\in \mathcal{D}}\sum_{i_{\ell}\in X}u(i_{\ell}, ~t_\textit{id})$.
\end{definition}

For example, $tu(t_{1})$ = \$26, and the transaction having the maximum utility is $t_6$ with $tu(t_{6})$. Moreover, the calculated transaction utility of each transaction is \{($t_{1}$: \$26), ($t_{2}$: \$14), ($t_{3}$: \$10), ($t_{4}$: \$18), ($t_{5}$: \$7), ($t_{6}$: \$27), ($t_{7}$: \$10)\}.

\begin{property}\label{GR}
	The \textit{TWU} complies with the property of downward closure. This means that if an itemset's $\textit{TWU(X)}$ is less than or equal to the utility threshold $xi$, any other itemsets containing $X$ are low-utility itemsets.
\end{property}

The \textit{TWU} values of all the items are determined by initially scanning the database $\mathcal{D}$. Property \ref{GR} states that when $\textit{TWU(X)}$ < $\xi$ for an item $X$, $X$ will be discarded to reduce the number of itemsets to be evaluated. To achieve the goal of fast mining, the paper uses a TWU ascending sort and processes each transaction in the transaction database according to that order as well. The transaction-weighted utility of all 1-itemsets is sorted as: $g$ (= \$28) $<$ $f$ (= \$45) $<$ $a$ (= \$49) $<$ $e$ (= \$50) $<$ $c$ (= \$55) $<$ $d$ (= \$63) $<$ $b$ (= \$92).

\begin{definition}
	\rm (Remaining utility) If there is an itemset $X$, and $X \subseteq t_\textit{id}$, the remaining items after $X$ in $t_\textit{id}$ is denoted as $t_\textit{id}/X$. The remaining utility \cite{liu2012mining} of itemset $X$ in the transaction $t_\textit{id}$ is the sum of the utilities of  $t_\textit{id}/X$ denoted as $ru(X,t_\textit{id})$, where $ru(X,t_\textit{id})$ = $\sum_{ i_{\ell} \in t_\textit{id}/X}u(i_{\ell}, ~t_\textit{id})$. Similarly, the remaining utility of $X$ in  $\mathcal{D}$ is denoted as $ru(X)$, where $ru(X)$ = $\sum_{X \subseteq t_\textit{id} \in\mathcal{D}}ru(X,t_\textit{id})$.
\end{definition}

In Table \ref{Tab:example}, the remaining utility of $\{bd\}$ in $t_1$ is defined as $ru(\{bd\},t_1)$ = $u(e,t_1)$ = \$2.

\begin{definition}
	\rm (Utility-list) In HUI-Miner \cite{liu2012mining}, the utility-list was introduced for the utility mining task. There are three essential elements ($t_\textit{id}$, \textit{iutil}, and \textit{rutil}) in the utility-list. The $t_\textit{id}$ is the identifier of a transaction that contains $X$. Element \textit{iutil} represents the utility of $X_i$ in $t_\textit{id}$. Similarly, \textit{rutil} is the remaining utility of $X_i$ in $t_\textit{id}$.
\end{definition}

For example, the utility-list of $\{c\}$ is \{$t_3$, \$6, \$4\}, \{$t_4$, \$8, \$4\} and \{$t_6$, \$6, \$18\}. The utility-list of $\{f\}$ is \{$t_4$, \$2, \$2\} and \{$t_6$, \$2, \$0\}. Therefore, the utility-list of $\{c, f\}$ is \{$t_4$, \$10, \$2\} and \{$t_6$, \$8, \$0\}.

\begin{definition}
	\rm (Target high-utility itemset) Let \textit{targetUtil} be the target utility threshold specified by a user and denoted as $\xi$. $X$ is known as a HUI for the user if its utility value is not smaller than  $\xi$. A target itemset  set by the user is called $T^\prime$. This itemset is called target since it is what the user wants to focus on. If $ u(X)$ $> \xi $, $T^\prime$ $\subseteq$ $X$, then $X$ is further recognized as one of the target high-utility itemsets (THUIs) \cite{miao2021targeted}.
\end{definition}

For example, a user wants to look for HUIs related to the target itemset $T^\prime$ = $\{b,d\}$. From Table \ref{Tab:hui}, we can search the HUIs with $T^\prime$. As a result, the set of THUIs is \{$\{f,c,d,b\}$, $\{e,d,b\}$, $\{c,d,b\}$, $\{d,b\}$\}.

\begin{definition}
	\rm (Top-$k$ target high-utility itemset) Inspired by other top-$k$ pattern mining tasks, we define the top-$k$ target high-utility itemsets (top-$k$ THUIs) as the $k$ targeted itemsets that have the highest utility values. The result itemsets will be less than $k$ items if the itemsets containing the target pattern in the database are relatively rare and the $k$ set by the user is relatively large. When there are more than $k$ patterns with the same utility, the results will also show only the $k$ itemsets in the order of mining.
\end{definition}

For example, if a user only wants to find the top-3 itemsets that satisfy the previous condition, then we set $k$ to 3. Obviously, the top-3 target HUIs are $\{f,c,d,b\}$ = \$27, $\{e,d,b\}$ = \$26, and $\{d,b\}$ = \$43.

\textbf{Problem statement}: Most utility mining algorithms are designed to identify itemsets with a utility equal to or greater than a fixed utility threshold. However, in reality, the user does not want to find all HUIs. For instance, a user may only want the top-$k$ HUIs containing a particular item or itemset. This paper calls the designed pattern mining problem as the targeted mining of top-$k$ high utility itemsets.

Unlike other HUIM problems, this paper needs to face the challenge of how to better deal with the search space under the combination of the target pattern and the top-$k$ pattern. In other words, it needs to apply a reasonable and efficient data structure to store the itemset information and a correct and fast pruning strategy to improve the mining efficiency. So far, the relevant definitions and concepts have been described, and the next section will introduce the details of the proposed TMKU algorithm.
\section{The TMKU Algorithm} \label{sec:algorithm}

\begin{figure}
	\centering
	\includegraphics[scale=0.35]{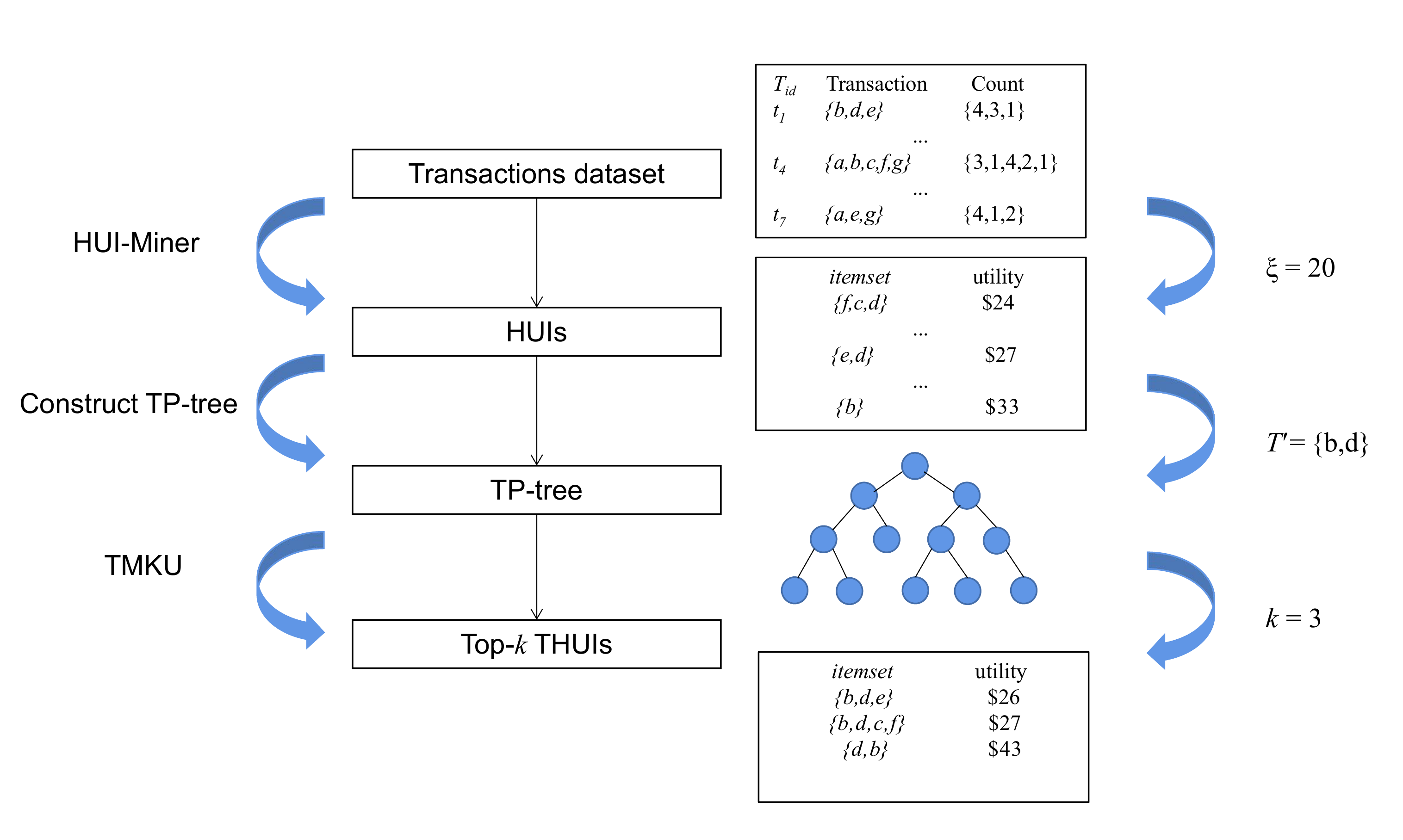} 
	\caption{The framework of TMKU}
	\label{img1} 
\end{figure}

The targeted mining of top-$k$ HUIs algorithm, abbreviated as TMKU, is presented at length in this section. Figure \ref{img1} is the framework diagram of the proposed TMKU algorithm, showing the basic steps of the whole algorithm. TMKU uses a trie structure as storage method similar to the TargetUM algorithm, and then utilizes the idea of filtering to dynamically change $\eta$ (the utility threshold) to select the desired top-$k$ THUIs. The TMKU algorithm enables a wide range of applications by adjusting the target pattern and $k$ to query different targets and the number of result itemsets in a short period of time. After presenting these concepts, we will explain how search space pruning is conducted as well as the processes of the TMKU algorithm in detail.

\subsection{Constructing the TP-tree and query mechanism}

For data analysis, the database must be scanned to process the information and obtain patterns that meet the user's requirements. When using the target-pattern tree as a data structure, the algorithm uses nodes to represent itemsets, and the pattern tree can be updated incrementally by inserting new nodes.

MEIT \cite{fournier2013meit} is an early study about this problem. It uses a compact node-compression mechanism to store information about transactions and answer queries about target itemsets, but it is not designed for utility mining. Later, the TargetUM algorithm was proposed for HUIM. The algorithm uses a trie tree to store HUIs from the whole database, and this was shown to reduce memory consumption. Therefore, the algorithm proposed in this paper also adopts the trie structure, which can reduce the overhead time for queries by using the common prefix of strings to improve efficiency. Then, after identifying all HUIs containing the target item or itemset, a map structure (TopKMap) is built to store the discovered target itemsets. The algorithm takes the utility value of the $k$-th itemset as the initial threshold, and constantly updates the TopKMap w.r.t. the minimum threshold until the top-$k$ target HUIs are discovered.

\begin{definition}
	\rm (Target pattern tree). There are seven elements describing each node in this tree. For a node $n$ (\textit{name}), its parent node (\textit{parent}) is recorded. In addition, the transaction weighted utilization (\textit{twu}), \textit{sumIu} which records the utility of the itemsets represented by the current node, and \textit{sumRu} which records the remaining utility of the itemsets as the current node, are the three elements related to the utility. When $n$ is the last item of a HUI, one variable (\textit{isEnd}) is set to true. If not, \textit{isEnd} is set to false. The last element indicates the link to the next node (\textit{link}) containing the same item.
\end{definition}

\begin{definition}
	\rm (Item header table). Each item has its own item header table. There is also a pointer to the first linked node corresponding to each item in the TP-tree. This table is useful for quickly locating the position of the item nodes.	
\end{definition}

\begin{figure}
	\centering
	\includegraphics[scale=0.35]{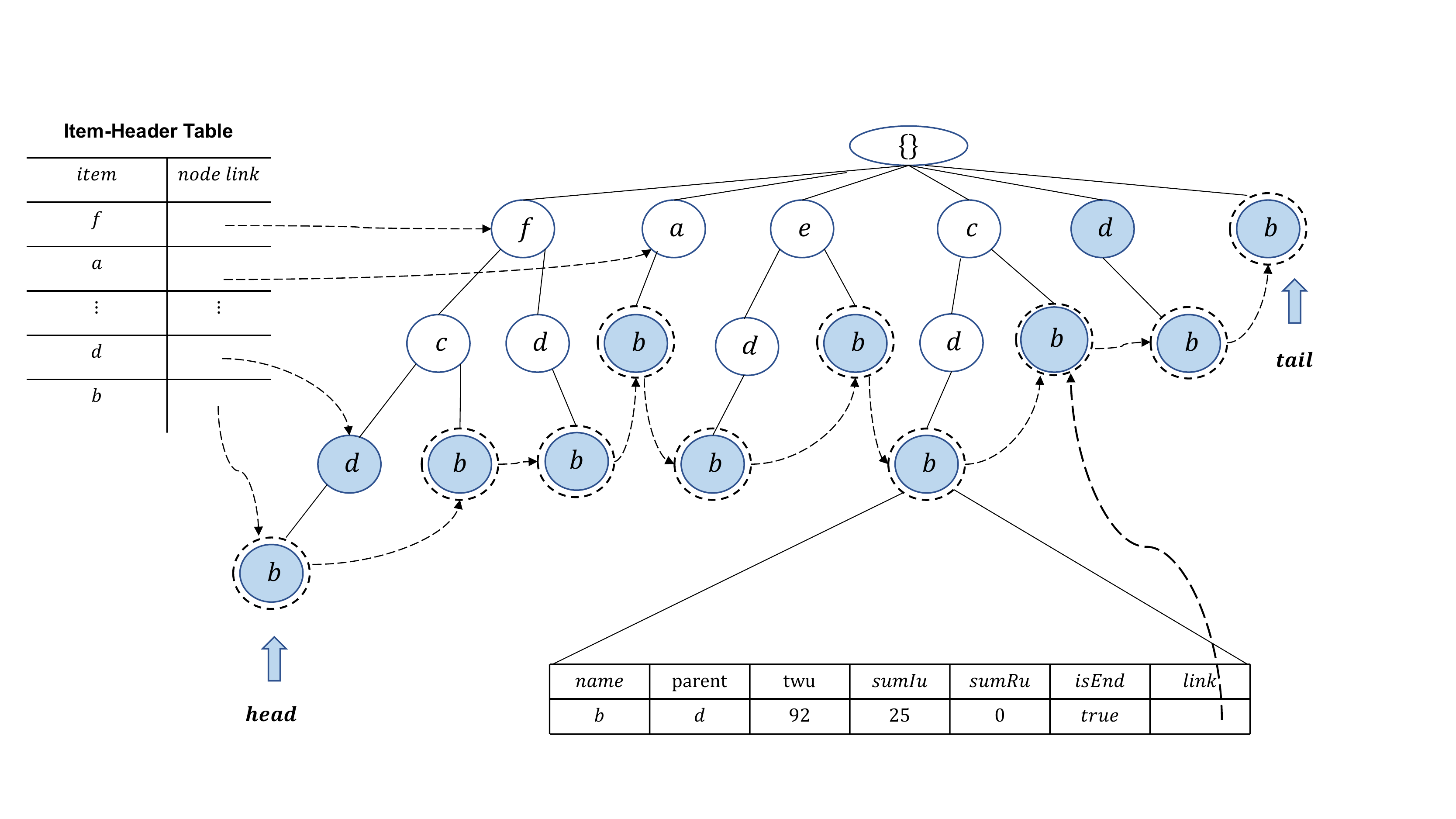} 
	\caption{The construction of TP-tree}
	\label{img2} 
\end{figure}

When the initial HUI \{$f$, $c$, $d$\} is found, the node $f$ is first added into this tree since it has the smallest \textit{TWU} value. Moreover, the element information of $f$ is recorded. Then, the algorithm creates the item header table of the $f$ node. The other nodes are then inserted in the same way. However, each time a node is inserted, the algorithm needs to verify whether a node with the same name already exists. If it already exists, the node is not created again. But the inserted itemset will share a common prefix with the already existing node item. When the second HUI \{$f$, $c$, $d$, $b$\} is inserted into the TP-tree, it shares the prefix ($f$, $c$, $d$) with the first HUI \{$f$, $c$, $d$\}. Then, node $b$ is created, and its information is recorded to update the item header table. Furthermore, the update of the item header table for item $b$ requires linking the new node. Subsequently, the last $b$ becomes the updated trailing node. Other HUIs that are stored in the TP-tree must also be inserted using the same process until the construction of the TP-tree is finished. The construction of the TP-tree is illustrated in Figure \ref{img2}.

\subsection{Targeted utility raising strategy}

As mentioned before, if only the value of $k$ is set by the user, the task of top-$k$ HUIM is to find out the top-$k$ HUIs. Since the utility threshold is not given by the user, effective utility raising strategies need to be proposed to efficiently find the top-$k$ HUIs. In this paper, the TMKU algorithm adopts two strategies (SUR and RIU) to achieve this goal.

\begin{strategy}
	\label{sur}
	\rm (SUR strategy). The SUR strategy \cite{zhang2021tkus} is a classic threshold raising strategy, used by the TMKU algorithm when storing the results in the TopKMap. The TopKMap can store $k$ items and their utility values dynamically. 
\end{strategy}

At first, the TopKMap is empty, and then when each THUI is found, the designed TMKU algorithm needs to consider whether to add it to the TopKMap or not. If the utility value of THUI is greater than $\eta$, it can be included in the TopKMap. However, when the number of itemsets stored in the TopKMap is greater than the $k$ value set by the user, some itemsets in the TopKMap need to be gradually deleted and updated. Those itemsets with low utility values will be replaced by itemsets with high utility values.

\begin{strategy}
	\label{riu}
	\rm (RIU strategy). The RIU raising strategy \cite{tseng2015efficient} is primarily based on the real utilities of items. If the number of THUIs is more than $k$ and the $k$-th highest utility value after sorting HUIs is still higher than $\eta$, the value of $\eta$ should be changed to the $k$-th highest value.
\end{strategy}

The RIU threshold raising strategy plays an integral role in the TMKU algorithm. The true utility of each item can be identified by scanning THUIs. To raise the utility threshold more quickly, TMKU uses the true utility of the first item as the reference value for $\eta$. Finally, combining with the SUR strategy, the correct top-$k$ THUIs can be quickly mined.

\subsection{Pruning strategies}

The TP-tree and item header table have been presented, which speeds up processing. Furthermore, some utility threshold-raising strategies are also adopted into the TMKU algorithm, which were listed in the previous subsection. Some pruning strategies used in TMKU are introduced in this subsection. They allow the algorithm to considerably reduce the search space and time.

\begin{strategy}
	\label{lab:one}
	\rm If the sum of utilities of $X_i$ $<$ $\xi$, then $X_i$ is not a THUI.	
\end{strategy}

It was shown in prior work that if \textit{TWU($X_i$)} < $\xi$, $u(X_i)$ is also less than the threshold since $u(X_i)$ must be smaller than \textit{TWU($X_i$)}. Therefore, it is not necessary to check any itemsets containing $X_i$. The utility and remaining utility of each item are recorded in its utility-list. If the sum of the utilities of $X_i$ (\textit{sumIu}) is smaller than $\xi$ defined by the user, then $X_i$ is not the HUI that we are looking for. Conversely, it is a HUI. Moreover, the utility-list also records the sum of the remaining utilities called \textit{sumRu}, which can be cleverly used to prune these utility values to improve search efficiency.

\begin{strategy}
	\label{lab:two}
	\rm In the TMKU algorithm, let $X^\prime$ and $\xi$ be the candidate and current minimum utility value, respectively. Suppose that the sum of \textit{sumIu} and \textit{sumRu} of $X^\prime$ is smaller than $\xi$. Then the supersets and extended itemsets of $X^\prime$ do not need to be checked.
\end{strategy}

In other words, the sum of \textit{sumIu} and \textit{sumRu} respects the downward closure property. Therefore, no extension representing $X^\prime$ can be the required top-$k$ THUI. This pruning strategy can prevent exploring several itemsets. It is worth noting that if the sum of \textit{sumIu} and \textit{sumRu} of $X^\prime$ is greater than $\xi$, then the extension of $X^\prime$ is a potential HUI and a further check is needed.

In addition, we can find that more and more branches may be generated during the construction of the TP-tree. However, the task is to discover all the top-$k$ HUIs that contain a target itemset, so not all high-utility itemsets must be searched. The TMKU algorithm needs to determine if the target itemset is included before inserting a node to reduce the running time. Therefore, TMKU uses the item header table to cut off the useless branches in advance and correctly and quickly identify the high-utility itemsets containing the target pattern. Then, we can use the item header table to quickly find the nodes in the target pattern tree.

\begin{strategy}
	\label{lab:three}
	\rm When $T^\prime$ is set as the target pattern, the TMKU algorithm sorts these items of the target itemset in \textit{TWU} ascending order. At this point, the item $i$ with the largest \textit{TWU} value is selected, and the position of $i$ in the TP-tree and its branches is determined using the item header table. A bottom-up approach is used to query whether this HUI contains the target pattern. When determining whether item $j$ in the target pattern is in the branch, we can compare the \textit{TWU} of the current node with the value of $\textit{TWU}(j)$. If the current node's \textit{TWU} is smaller than $\textit{TWU}(j)$, the branch does not contain the target pattern and can be discarded directly. If the current node's \textit{TWU} is greater than $\textit{TWU}(j)$, further exploration is required. If equal, it is also necessary to check whether the current node's node name is $j$ to avoid replacing the same \textit{TWU} value items.
\end{strategy}

For example, supposing that \{$f$, $c$, $d$, $b$\} is a HUI, and \{$a$, $b$\} is denoted as $T^\prime$. First, we find that item $b$ whose \textit{TWU} is the highest in the target pattern. Then, we need to continue checking whether $a$ is in the HUI, and then compare the \textit{TWU} values of other items in the HUI with the \textit{TWU} of $a$ in turn. It is found that the \textit{TWU} of both $d$ and $c$ is greater than $a$, but the \textit{TWU} of $f$ is less than $a$. According to the above analysis, it is known that \{$a$, $b$\} is not in the HUI and the search is stopped.

\subsection{Proposed algorithm}

The TMKU task will be summarized as follows, based on the above introduction. To discover those itemsets containing the target itemset, Algorithm \ref{algo:THUI} focuses on accomplishing this task by constructing the TP-tree and the item header table. Algorithm \ref{algo:TMKU} introduces how to search the TP-tree to find the THUIs followed by mining the top-$k$ itemsets, and through some utility threshold raising strategies, users can get the desired results as soon as possible.

%%%%%%%%%%%%%%%%%   Construct TP-tree procedure   %%%%%%%%%%%%%%%%%%%
\begin{algorithm}[h]
	\small
	\caption{The construction procedure}
	\label{algo:THUI}
	\LinesNumbered
		\KwIn{$X^\prime$: the prefix of HUI; \textit{IUs}: the utility-list of $X^\prime$; \textit{RUs}: the remaining utility-list of $X^\prime$; \textit{TUs}: the \textit{TWU} list of $X^\prime$; $x$: the last item of HUI; $\alpha$: the utility of $x$; $\beta$: the remaining utility of $x$; $\theta$: the \textit{TWU} of $x$; \textit{mapItemNode} store the header node of each item-header table; \textit{mapItemLastNode}: store the tail nodes of each item-header table; $\xi$: the target utility threshold.}
		\KwOut{\textit{TP-tree}; item header table.}
		
        initialize \textit{TP-tree}, \textit{currentNode} = \textit{null}, \textit{parentNode} = \textit{null}, \textit{listNodes} = \textit{root.childs}\;
        insert a HUI into the TP-tree\;
        \For{\rm each item $e$ $\in$ $X^\prime$}{
           search the position of $e$\;
           \If{\textit{currentNode} == \textit{null}}{
            create a new node in the TP-tree and construct item header table of $e$;
       }
       update \textit{parentNode}, \textit{listNodes}\;
      }
      \textit{currentNode} = $x$\;
      \eIf{\textit{currentNode} != \textit{null}}{
      {update} \textit{currentNode.sumIu}, \textit{currentNode.sumRu}, \textit{currentNode.TWU}\;
  }{ 
    construct new item header table of $x$\;
}
\textit{currentNode.isEnd} = \textit{true};
\end{algorithm}

%%%%%%%%%%%%%%%%%%   TMKU procedure  %%%%%%%%%%%%%%%%%%%

\begin{algorithm}[h]
	\small
	\caption{The discovery procedure}
    \label{algo:TMKU}
	\LinesNumbered
	\KwIn{$k$: the number of THUIs; $T^\prime$: the itemset obtain target; $\xi$: the target utility threshold; $\eta$: the minimum utility threshold.}
	\KwOut{top-$k$ THUIs.}  
	
	\textit{posToMatch} = |$T^\prime$|, \textit{node} = the target item of last item\;
	\While{\textit{node} != \textit{null}}{
	    \If{\rm \textit{node.sumIu} + \textit{node.sumRu} $\geq$ $\xi$}{ 
           \textit{node.name} = the current \textit{HUI} $X$\;
           \textit{posToMatch} decrease 1, {update} \textit{currentNode} = \textit{node.parent}\;
           
           \While{\textit{currentNode} != \textit{null}}{
           	  \If{\textit{posToMatch} $\geq$ 0}{
           	      a new item $y$ = $T^\prime$[\textit{posToMatch}]\;
           	      \If{\textit{currentNode.TWU} < \textit{y.TWU}}{
           	      {break}\;
           	    }
           	      \If{\textit{currentNode.TWU} == \textit{y.TWU} \textit{AND}  \textit{currentNode} == $y$}{
           	      	
                  \textit{posToMatch} decrease 1\;
              }
             }
              {update} $x$ and currentNode\;
          }
           	\If{\textit{posToMatch} == -1}{
           		
           	\If{\textit{node.sumRu} $\geq$ $\xi$ \textit{AND} \textit{currentNode.isEnd} = $true$}{
              a THUI has been discovered\;
           }
           recursively explore all suffix nodes of $X$\;
         }
		}
   }
	
   initialize TopKMap = $\emptyset$\;
   \For{\rm  each entry itemset $t$ in THUIs}{
    	\eIf{\textit{t.utility} $\geq$ $\eta$}{
            TopKMap.add(t)\;}{ 
            skip $t$ from TargetUM\;}
        \If{\rm the number of \rm{TopKMap} $>$ $k$}{
          {update} $\eta$ as the $k$-th highest utility in TopKMap\;
          {update} TopKMap and keep $k$ target itemsets with the highest utilities in TopKMap\;
	  }
   	    }
    \textbf{return} top-$k$ THUIs
\end{algorithm}

\textbf{(1) The construction of TMKU}: this procedure can be divided into two parts: the construction of the TP-tree and the processing of the item header table. The construction of a TP-tree is the basis of the TMKU algorithm. There are several indispensable parameters need to be introduced, including its last item $x$, the list of three utilities (\textit{IUs}, \textit{Rus}, \textit{TUs}), and the three utility values of $x$. The \textit{mapItemNode} is available to store the header node of each item header table. During the construction of the TP-tree, the node used to store the last item in each itemset is called the tail node, and the tail node is continuously updated. Besides, the \textit{mapItemLastNode} is required to store the tail nodes of each item header table. First, the algorithm initializes the TP-tree, i.e., sets the initial values for the several parameters (line 1). Based on the HUIs obtained by applying HUI-Miner, once a new HUI is found, we should consider inserting it into the TP-tree (line 2). For each item $n$ in $X^\prime$ (line 3), the position of $n$ in this TP-tree can be found easily (line 4). When there are no $n$ nodes in the TP-tree, the TMKU algorithm creates a new node and stores its information in the item header table (lines 5--6). After that, the values of nodes \textit{parentNode} and \textit{listNodes} are updated (line 8). When all the items of $X^\prime$ have been processed, the TP-tree of $X^\prime$ is completely built (lines 3--8). Then, while $x$ previously existed (found by the binary search method) (line 10), the algorithm will update the information value of the current node (lines 11 to 12). And if it does not exist, an item header table of $x$ is created (line 14). Finally, the value of the \textit{currentNode.isEnd} variable is set to true (line 16).

\textbf{(2) The discovery of TMKU}: The procedure for processing the target query is a depth-first search. First, the procedure determines whether the received prefix appears. Therefore, the node is denoted the last item of $T^\prime$, and then the \textit{posToMatch} parameter is initialized to record the current two items (line 1). Then, each item of $T^\prime$ is traversed. If the sum of \textit{sumLu} and \textit{sumRu} is not less than $\xi$, according to Strategy \ref{lab:two}, $X$ may be a THUI, and further comparison with other items can be performed (lines 2--25). Moreover, according to Strategy \ref{lab:three}, if the \textit{TWU(\textit{currentNode})} is smaller than the current item of $T^\prime$, it can be discarded directly, and the comparison is aborted (lines 9--10). When the value of \textit{posToMatch} decreases to -1, one of the target itemsets is found (lines 19--20). At this point, the algorithm uses a recursive method to output this THUI. First, all suffix nodes of $X$ are explored, and from each node, the extension nodes are obtained (line 22). Then, the THUIs are stored by filtering out the high-utility itemsets that do not contain the target itemset. Finally, if the remaining suffix nodes meet the condition of Strategy \ref{lab:two}, they need to be explored.

After that, the algorithm obtains the target high-utility itemsets. Then, the task of the algorithm is to quickly mine the top-$k$ THUI. First, TopKMap is initialized to the empty set and the utility value of the first THUI is used to initialize $\eta$ (line 26). Then, the algorithm iterates through all the itemsets in the THUIs and if the utility value of the itemset is higher than $\eta$, the itemset is added to the TopKMap (lines 27--29). Otherwise, the THUI is removed (line 31). When the size of the TopKMap is over $k$ (line 33), $\eta$ is updated to the $k$-th highest utility in the TopKMap (line 34). Moreover, the algorithm updates the TopKMap and keeps the $k$ target itemsets with the highest utilities in the TopKMap (line 35). Finally, all the top-$k$ THUIs can be explored and output successfully (line 38).

\section{Performance Evaluation}  \label{sec:experiment}

The TMKU algorithm was evaluated by a series of tests on various types of datasets. By comparing the results, the correctness and efficiency of the TMKU algorithm are proved. The TMKU algorithm, as previously noted, is the first to discover the top-$k$ HUIs that include a target pattern. All the current top-$k$ HUIM algorithms cannot successfully tackle the challenge of considering the user's needs. As a result, none of the existing methods can be compared to analyze and evaluate the proposed TMKU algorithm.

On the one hand, to determine the accuracy of the algorithm, we compared the results of the TMKU algorithm with those of the TargetUM algorithm after post-processing on various datasets. On the other hand, we conducted various experiments on six datasets to determine the efficiency of TMKU in a comprehensive manner and evaluated three variants using different pruning strategies. It is worth noting that TMKU$_{V1}$ only uses the Strategy \ref{lab:one} and Strategy \ref{lab:three}, and TMKU$_{V2}$ uses the Strategy \ref{lab:two} and Strategy \ref{lab:three}. When all pruning strategies are used, TMKU is obtained. As a result, three versions are used to evaluate the efficiency of TMKU.

\subsection{Experimental setup}

All the algorithms were programmed in Java. The PC used for experiments is equipped with 8.0 GB of RAM, 64-bit Windows 10, and an AMD Ryzen 5 3600 CPU.

To assess the effectiveness and scalability of the TMKU algorithm, experiments were carried out on a variety of datasets, including real and synthetic databases. The SPMF data mining library is the source of all the datasets\footnote{http://www.philippe-fournier-viger.com/spmf/}. These datasets have different characteristics, which makes it beneficial to observe the performance of the algorithm in different situations. Table \ref{table:data} lists all the features of the experimental datasets. A brief description is given below.

\begin{itemize}
	\item \textbf{Retail} is a real-life dataset with many items in each transaction. Transactions come from a store in Belgium and the utility values are synthetic.
	
	\item \textbf{Foodmart} is a transaction dataset from a grocery store where items are sparse and transactions are short.
	
	\item \textbf{Chainstore} is a transaction dataset containing over a million transactions from a chain store in California.
	
	\item \textbf{Ecommerce} is a transactional dataset that contains all the transactions of an online retail store and has real utility values.
	
	\item \textbf{Mushroom} is a dense and real-world database with over 8,000 transactions, where the \textit{MaxLen} and \textit{AvgLen} are the same, and utility values are synthetic.
	
	\item \textbf{T10I4D100K} is synthetic data generated by the IBM Data Generators, which contains 8,124 transactions.
\end{itemize} 

\begin{table}[h]   
	\begin{center}   
		\caption{The characteristics of datasets}  
		\label{table:data} 
		\begin{tabular}{|m{2.8cm}<{\centering}|m{2.8cm}<{\centering}|m{2.8cm}<{\centering}|m{2.8cm}<{\centering}| m{2.8cm}<{\centering}|}   
			\hline   \textbf{Dataset} & \textbf{Size} & \textbf{Trans} & \textbf{MaxLen} & \textbf{AvgLen}\\   
			\hline  foodmart & 176 & 4141  & 4.4 & 14 \\
			
			\hline  retail & 1845 & 24735 & 10.3 & 74\\ 
			
			\hline  ecommerce & 1968 & 14976 & 11.6 & 29  \\  
			
			\hline  chainstore & 81102 & 1112950 & 7.2 & 170  \\ 
			
			\hline  mushroom & 1309 & 8124 & 23 & 23  \\
			
			\hline  T10I4D100K & 8267 & 100000 & 29 & 10.1  \\  
			\hline   
		\end{tabular}   
	\end{center}   
\end{table}

\subsection{Experiments on itemset}

The output of the TMKU algorithm is the top-$k$ THUIs. To evaluate the accuracy of the results, an experiment was done to compare the output of TMKU with that of TargetUM after post-processing the results. Here, $u_1$ represents the lowest utility in TMKU's result set, and $u_2$ denotes the same for the post-processed results from the TargetUM algorithm. We tested various $k$ values for different datasets due to the different numbers and features of each dataset.

The results mainly demonstrate the effect of changing the $k$-value by comparing the minimum utility in the dataset. Table \ref{table:right} indicates that $u_1$ and $u_2$ have the same value, which becomes smaller and smaller as $k$ is increased. Besides, it is found that the results are the same for both algorithms. For example, we randomly select watermelon as the user-selected target pattern for the retail database, so $T^\prime$ is set as \{976\} which represents the watermelon. Then, the experiment compares the minimum utility value for different $k$ values and $u_1$ and $u_2$.

It is found that when $k$ is increased from 10 to 110, the minimum utility value decreases from 4,325 to 1,389. TMKU described in this paper can successfully discover the top-$k$ THUIs, depending on the aforementioned analysis. Table \ref{table:candidate} displays the number of candidates in the six databases for different $k$ values. For example, we set $T^\prime$ = \{110\} for the mushroom database, which indicates the flavor of a certain type of mushroom. When $k$ is increased from 1,000 to 6,000, the number of candidate itemsets increases from 16,978 to 65,512. As the $k$-value increases, obviously, the number of candidate itemsets increases as well. This is in line with the expected results of the top-$k$ algorithm.

\begin{table}[h]
	\centering
	\caption{The minimum utility of TMKU with different $k$ values}
	\label{table:right}
	\begin{tabular}{c|c|lllllll}
		
		\hline \hline
		\multirow{2}{*}{\textbf{Dataset}} & \multirow{2}{*}{\textbf{u}} &
		\multicolumn{6}{c}{\textbf{\# minimum utility by varying $k$}} \\ \cline{3-8}
		& &$k_1$ & $k_2$ & $k_3$ & $k_4$ & $k_5$ & $k_6$ \\ \hline
		chainstore & $u_1$ & 1643,851 & 697,152 & 488,009 & 407,079 & 348,233 & 310,306\\
		\{16967\} & $u_2$ & 1643,851 & 697,152 & 488,009 & 407,079 & 348,233 & 310,306\\ \hline
		ecommerce & $u_1$ & 976,618 & 954,154 & 940,042 & 929,362 & 920,962 & 913,834 \\
		\{150561222\} & $u_2$ & 976,618 & 954,154 & 940,042 & 929,362 & 920,962 & 913,834 \\ \hline
		retail & $u_1$ & 1,381  & 1,379 & 1,377 & 1,377 & 1,376 & 1,375 \\
		\{976\} & $u_2$ & 1,381  & 1,379 & 1,377 & 1,377 & 1,376 & 1,375 \\ \hline
		foodmart& $u_1$ & 6,266 & 4,319 & 5,541 & 6,530 & 7,420 & 7,978 \\
		\{1340\} & $u_2$ & 6,266 & 4,319 & 5,541 & 6,530 & 7,420 & 7,978 \\\hline
		mushroom & $u_1$ & 449,193 & 358,323 & 256,158 & 125,079 & 93,238 & 73,171 \\
		\{110\} & $u_2$ & 449,193 & 358,323 & 256,158 & 125,079 & 93,238 & 73,171 \\\hline
		T10I4D100K & $u_1$ & 93,238 & 39,273 & 30,843 & 27,984 & 26,228 & 25,716 \\
		\{71\} & $u_2$ & 93,238 & 39,273 & 30,843 & 27,984 & 26,228 & 25,716 \\
		 \hline 
		\hline
	\end{tabular}
\end{table}

\begin{table}[h]
	\centering
	\caption{The number of candidates of TMKU with different $k$ values}
	\label{table:candidate}
	\begin{tabular}{c|c|c|c|c|c|l}
		
		\hline \hline
		\multirow{1}{*}{\textbf{Dataset}}  &
		\multicolumn{6}{c}{\textbf{\# the number of candidates by varying $k$}} \\ \cline{2-7}
		 &$k_1$ & $k_2$ & $k_3$ & $k_4$ & $k_5$ & $k_6$ \\ \hline
		chainstore \{16967\}  & 75 & 193 & 233 & 254 & 261 & 267 \\
		  \hline
		ecommerce \{150561222\}  & 43,509 & 84,230 & 123,914 & 160,464 & 196,604 & 231,791 \\ \hline
		retail \{976\} & 4,430  & 7,771 & 10,557 & 13,064 & 15,193 & 17,242 \\ \hline
		foodmart \{1340\} & 2,653 & 4,319 & 5,541 & 6,530 & 7,420 & 7,978 \\\hline
		mushroom \{110\} & 461 & 676 & 997 & 1248 & 1345 & 1456 \\\hline
		T10I4D100K \{71\} & 7,761 & 13,678 & 18,639 & 23,178 & 27,431 & 31,569 \\\hline 
		\hline
	\end{tabular}
\end{table}

\subsection{Experiments on runtime}

\begin{figure}
	\centering
	\includegraphics[width=1\textwidth]{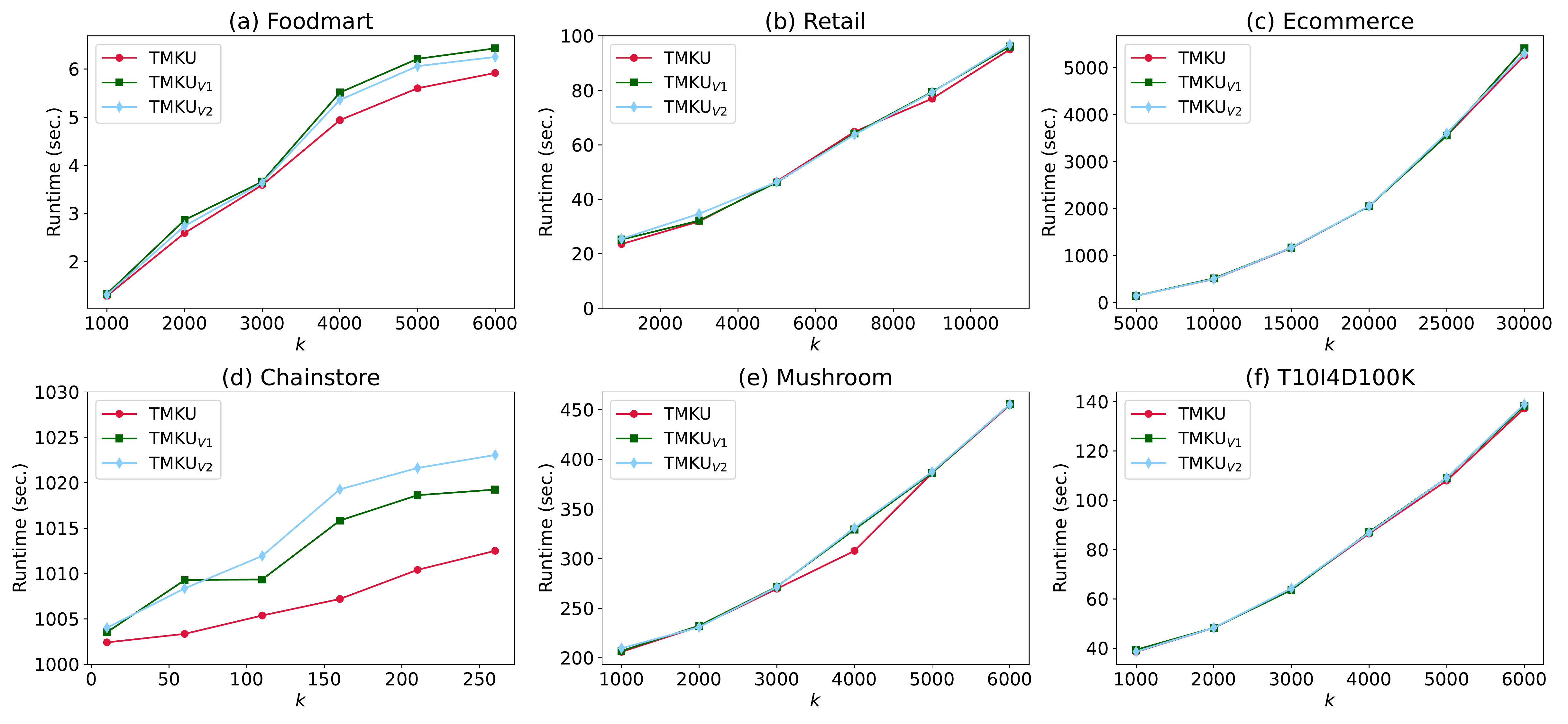} 
	\caption{Runtime under varied $k$. (a) foodmart ($T^\prime$ = \{1340\}). (b) retail ($T^\prime$ = \{1370\}). (c) Ecommerce ($T^\prime$ = \{150561222\}). (d) Chainstore ($T^\prime$ = \{16967\}). (e) Mushroom ($T^\prime$ = \{110\}). (f) T10I4D100K ($T^\prime$ = \{71\})}. 
	\label{runtime} 
\end{figure}

Experiments were also done to compare the running times of several versions of the algorithm as $k$ is changed. Figure \ref{runtime} shows the trend of the execution time of the algorithm for various $k$ values on the six datasets. It is worth noting that the top-$k$ algorithm has only one parameter, which could result in varying performances. In general, for the TMKU task, it is tough to determine an appropriate value of $k$ to find the optimal threshold that provides the best performance. As a result, the $k$-value in the TMKU algorithm is totally determined by the user's needs in most scenarios. This study is of great significance in targeted mining to consider the subjective needs of users and provide them with a positive experience.

The performance of the three algorithms on distinct datasets was recorded in the experiment to have a fair evaluation, which is depicted in Figure \ref{runtime}. The results reveal that when all the strategies are applied, the running time of the TMKU algorithm is shorter than that of other versions. The entire runtime increases as $k$ grows. Due to the limitations of the TP-tree, the runtime of the three versions does not reflect significant differences. However, we can observe that this difference can be expressed in Figure \ref{runtime} (a) and (d). Both the foodmart and chainstore datasets are sparse, with the chainstore dataset being particularly sparse. In different $k$ settings, therefore, the running time difference of three variants can be clearly observed. The distinction between the retail database and the e-commerce database is not immediately obvious.

In summary, the algorithm performs better on sparse datasets with pruning strategies. For example, in the chainstore dataset, when $k$ = 210, the proposed algorithm with all pruning strategies clearly takes less time than TMKU$_{V1}$ and TMKU$_{V2}$. TMKU takes 1,010 seconds to discover all the top-$k$ THUIs, while TMKU$_{V1}$ and TMKU$_{V2}$ need 1,019 seconds and 1,025 seconds to finish this task, respectively. This result demonstrates that all the strategies are effective for mining the top-$k$ THUIs. Furthermore, Strategy \ref{lab:one} and Strategy \ref{lab:two} work on the prefix and suffix of itemset, separately, which are crucial in speeding up the query. Strategy \ref{lab:three} has a pruning effect for the case when the target pattern $|$$T^\prime$$| >$ 1, as mentioned in the TargetUM task.

\subsection{Experiments on memory}

\begin{figure}
	\centering
	\includegraphics[width=1\textwidth]{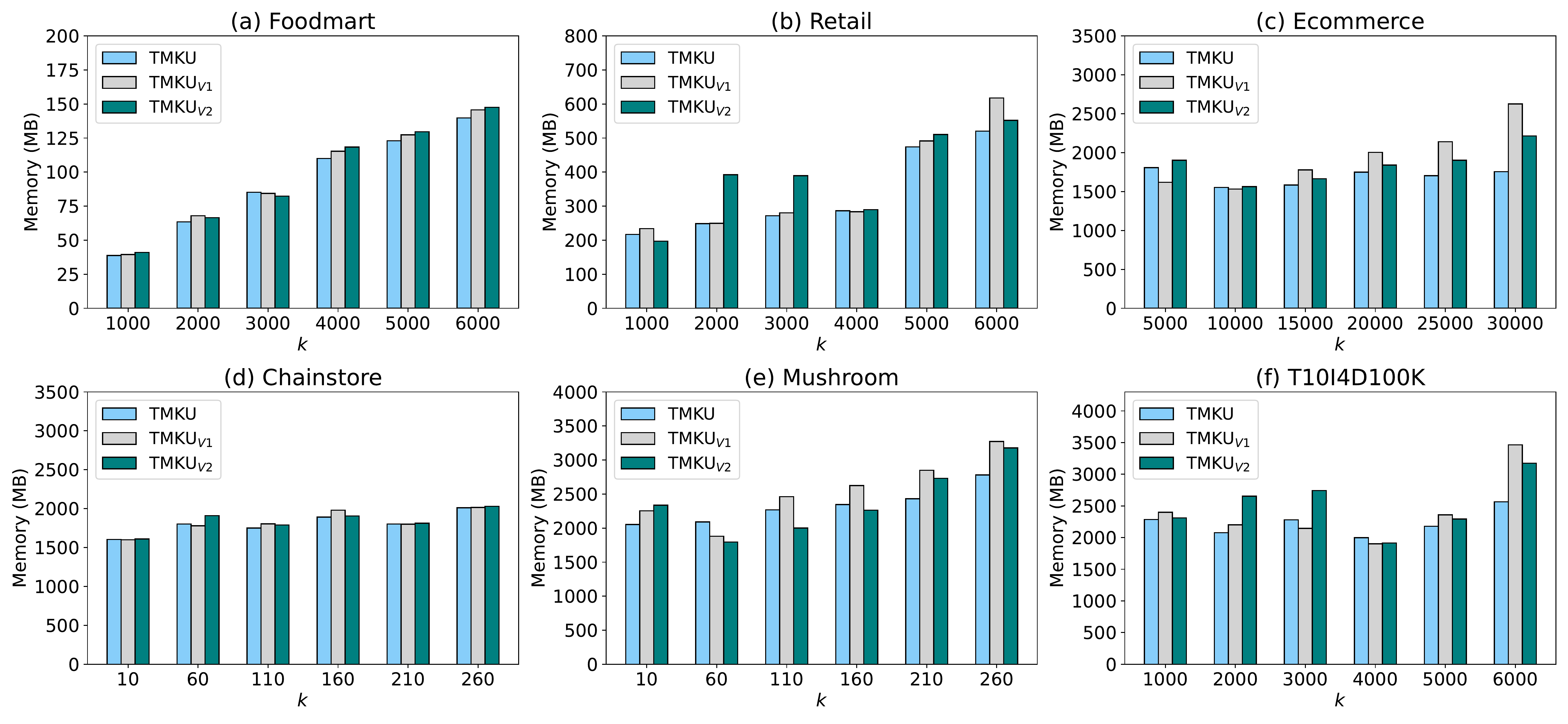} 
	\caption{Memory for varied $k$ values. (a) foodmart ($T^\prime$ = \{1340\}). (b) retail ($T^\prime$ = \{1370\}). (c) Ecommerce ($T^\prime$ = \{150561222\}). (d) Chainstore ($T^\prime$ = \{16967\}). (e) Mushroom ($T^\prime$ = \{110\}).  (f) T10I4D100K ($T^\prime$ = \{71\})}.
	\label{memory} 
\end{figure}

Additional tests were done to check how much memory is used by each version of the algorithm when the $k$ parameter is varied, and Figure \ref{memory} shows the detailed results. We can clearly see that the algorithm that uses all pruning strategies consumes less memory than the other variants. This is due to the fact that some memory is consumed during the first phase of the TP-tree construction, but the memory impact of these pruning strategies is not significant. In addition, the overall memory usage of TMKU increases as $k$ increases in all three versions, eventually reaching equilibrium. Since the top-$k$ THUIs are mined by using the Strategy \ref{sur} and Strategy \ref{riu} to save the items and real utility values of items, more storage space is necessary in the second step of the algorithm. However, under certain conditions, the algorithm can fluctuate slightly in the amount of memory consumed while running. For example, for the foodmart dataset, as $k$ is gradually increased from 1,000 to 6,000, the memory usage is also gradually increasing. However, when $k$ is set to 4,000 in the T10I4D100K database, the memory usage is slightly less than before. In addition, the memory usage is more stable for datasets like chainstore where the number of transactions is high, and the utility values are concentrated. In general, the TMKU algorithm has pretty good capability in terms of memory usage.

\subsection{Experiments on scalability}

\begin{figure}
	\centering
	\includegraphics[scale=0.3]{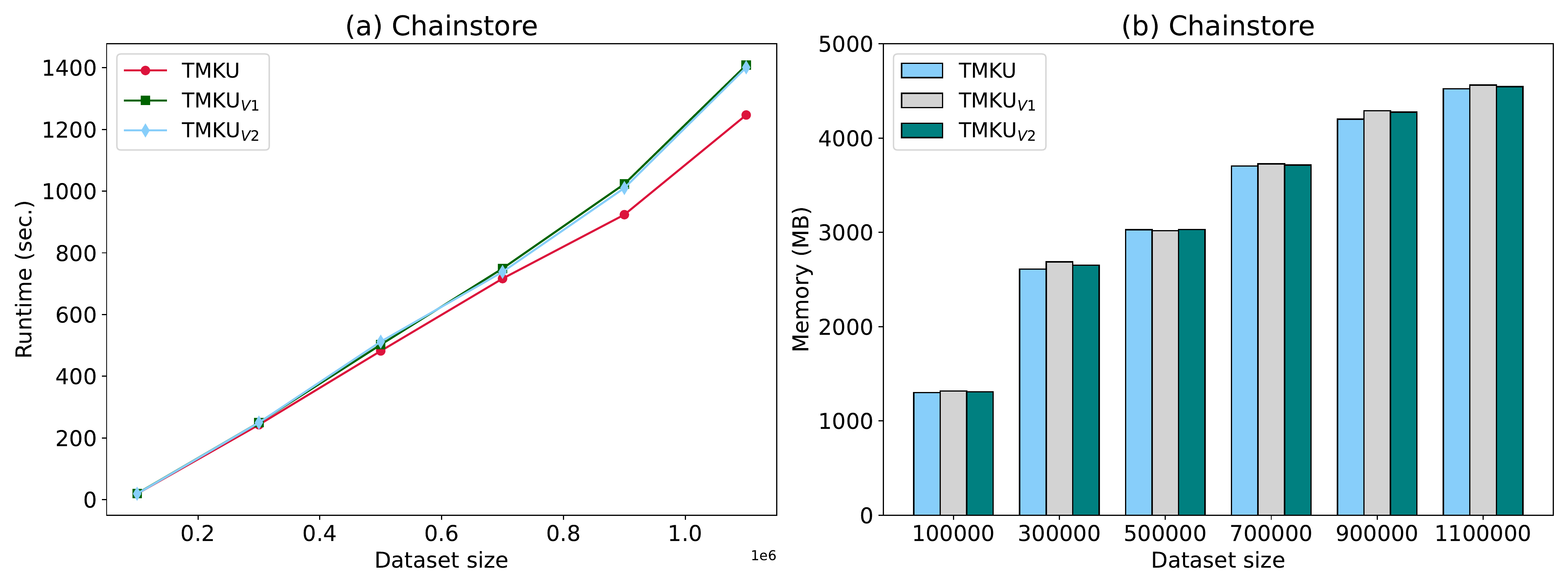} 
	\caption{The runtime and memory scalability of variants on Chainstore ($T^\prime$ = \{16967\}, $k$ = 50)}.
	\label{extend} 
\end{figure}

Finally, to better assess the performance of the TMKU algorithm, the scalability of TMKU was tested on the chainstore dataset. The size of the chainstore dataset was increased linearly from 100K to 1,100K. The experimental results in terms of execution time and memory for the three variants of the TMKU algorithm are shown in Figure \ref{extend} for the case of $k$ = 50. Overall, the TMKU algorithm performs the best among the three algorithms. It can be clearly seen that as the number of transactions in the database increases, the execution time increases as well. For instance, when the chainstore database size grows from 300K to 900K, the execution time of TMKU increases from 242.86s to 923.79s. When compared to TMKU$_{V1}$ and TMKU$_{V2}$, TMKU takes the least time. As far as memory usage is concerned, we can observe that memory consumption in general also increases as the data set size increases. However, the performance of strategies adopted by TMKU is subject to fluctuations due to the instability of memory consumption. When the dataset size is 500K, for example, TMKU's memory consumption differs very little from that of TMKU$_{V1}$ and TMKU$_{V2}$. Figure \ref{extend} clearly shows that the runtime and memory consumption increase linearly with the size of the dataset. Besides, the TMKU algorithm is less time-consuming and space-consuming than the other two variants, so it has good scalability.

\section{Conclusion} \label{sec:conclusion}

Until now, no algorithm has been proposed to mine the targeted top-$k$ high utility itemsets. In this research, the TMKU algorithm is proposed for this task. The algorithm utilizes several data structures, namely the utility list, TP-tree, TopKMap, and two threshold-raising strategies (SUR and RIU) to achieve the task. In addition, several pruning strategies are applied in the TMKU algorithm to lower the execution time. Under this task, users can determine $k$ and $T^\prime$ according to their needs. After testing the TMKU algorithm on six datasets, it was found that TMKU is able to create the TP-tree completely, raise the threshold quickly, and obtain the top-$k$ THUIs efficiently. Moreover, TMKU has short processing time, stable memory usage, and excellent scalability. The TMKU algorithm presents a solution to a new problem in the field of data mining, and in the future, we will work on designing more advanced algorithms to explore this promising topic. Furthermore, to explore more applications of TMKU and extend it to more scenarios, such as using it in a distributed environment. Future research could also take into account visualization of mining results to boost interpretability.

\section*{Acknowledgment}

This research was supported in part by the National Natural Science Foundation of China (Nos. 62002136, 62272196, 61472049, and 61572225), Natural Science Foundation of Guangdong Province (No. 2022A1515011861), Guangzhou Basic and Applied Basic Research Foundation (No. 202102020277), the Young Scholar Program of Pazhou Lab (No. PZL2021KF0023), Engineering Research Center of Trustworthy AI, Ministry of Education (Jinan University), and Guangdong Key Laboratory for Data Security and Privacy Preserving..
%%%%%%%%%%%%%%%%%%%%%%%%%%%%%%%%%%%%%%%

\printcredits

\bibliographystyle{cas-model2-names}

\bibliography{TMKU.bib}

\end{document}